\documentstyle[aps,preprint]{revtex}
\begin{document}

\begin{center}
{\large\bf Cooperativity in a trading model with memory and production}
\vspace{1.0cm}

{\large R.\ Donangelo$^a$ and K.\ Sneppen$^b$}
\vspace{1.0cm}

$^a$ Instituto de F\'\i sica, Universidade Federal do Rio de Janeiro,\\
C.P. 68528, 21945-970 Rio de Janeiro, Brazil.
\vspace{0.3cm}

$^b$ Department of Physics, Norwegian University of Science and Technology,\\
N-7491, Trondheim, Norway.
\vspace{0.3cm}

\end{center}
\vspace{1.5cm}

\begin{small}
\noindent
We consider in a market model the cooperative emergence of value 
due to a positive feedback between perception of needs and demand.
Here we consider also a negative feedback from production
of the traded products, and find that this cooperativity is
robust, provided that the production rate is slow.
Cooperativity is found to be critically linked to the ability
to minimize the overall need, and thus disappears when the agents are
poor, when the production rate is large or when there is little trade.
We further observe that a cooperative economy may self-organize to 
compensate for an eventual slow production rate of certain products,
so that these products are found in sizeable stocks.
This differs qualitatively from an economy where cooperativity did not
develop, in which case no product has a stock larger than what
its bare production rate justifies.
We also find that these results are robust in relation to the
spatial restriction of the agents.
\end{small}

\vspace{1.0cm}
\vfill\eject
\section {Introduction.}

Recently physicists have begun applying Statistical Mechanics and other
tools employed to study complex systems, in an effort to understand the
properties of financial and commodities markets \cite{Farmer,BPS,Minority,CZ97,B01}.
A particular striking feature of real markets is their ability to develop
collective modes, including features such as bubbles and collective crashes
of a whole stock market.
Such features are not possible within the framework of equilibrium and only
negative feedbacks, but may be associated to the positive social feedback
mechanisms that must be part of building any society.
Positive feedbacks are considered in some economic literature, most notably
in H. Simon's article on emergence of Zipf laws in various social settings
\cite{Simon}, and in W.B. Arthur's discussion \cite{Arthur} on the 
initial development of frozen states in manufacturing activities.
The existence of such factors containing self-amplifying mechanisms 
brings about the interplay between these and the classical negative feedback 
mechanisms between the amount of any product and its value.
This calls for new model building.

In the present paper we show how collective modes in trading markets may be
reproduced by means of a very schematic model, previously discussed
in \cite{DS00}, and conceptually related to \cite{Yasutomi}.
We have already exemplified how the model may explain the appearance of money
in a primitive economy, and how the concept of demand may be associated
to the memory of past transactions by the different agents.
Other properties of the model are presented in Refs.~\cite{DHSS00} and \cite{SR01}.
Here we study how production and consumption integrate into the market model.

We have organized this work as follows. In Section 2 we provide a short
description of the model and the most relevant features found in previous
studies. In Section 3 production and consumption are introduced and their
effects on demand, stock size and monetary value for given products are
studied. In the concluding Section 4 we discuss these results and make
suggestions for future work. 

\section {Description of the model.}

The market we consider consists of $N_{ag}$ agents
and $N_{pr}$ different products.
Initially we give $N_{unit}$ units of the products to each agent.
The number $N_{unit}$ is fixed, but the products are chosen at
random, so the individuals are not in exactly the same situation.
At each timestep we select two agents at random and let them attempt
to perform a trade between each other.
In order to perform a trade, each agent presents the other a list
with the goods he is interested to obtain.

In \cite{DS00} this was done in two steps. 
The trade started by comparing the list of goods that each agent
lacks and therefore would like to get from the other agent in
exchange for goods it has in stock.
Therefore, the model first considered the simple ``need'' based 
exchange procedure: when each of the agents had products that the 
other needed, then one of these products, chosen at random, was exchanged.
In case such a ``need'' based exchange were not possible they considered
the ``greed'' exchange procedure: one or both of the agents would
accept goods which they do not lacked, but considered useful for
future exchanges.

In order to determine the usefulness of a product, each agent $i$
kept a record of the last requests for goods it received in
encounters with other agents.
This memory is finite, having a length of $N_{mem} (i)$ positions,
each of which registers a product that was requested.
As the memory gets filled, the record of old transactions is lost.
Agents accepted products they already have in stock with a probability
based on their memory record.
The chance that agent $i$ accepting such a good $j$ was taken to be
proportional to the number of times $T_{ij}$ that good $j$ appears on
the memory list of agent $i$:
\begin{equation}
p_{ij} \;=\; \frac{T_{ij}}{N_{mem}(i)}\;.
\label{Prob}
\end{equation}

Here we simplify the procedure by making a single list containing
all the products the agent needs and all those he would accept in
exchange. A product is included in this list with probability given
by Eq.~(\ref{Prob}). We have verified that the properties of the
model, described in \cite{DS00} are not modified with this
simplification. These general properties are:
\begin{itemize}
\item{}
After a relatively short equilibration time ``greed'' transactions 
dominate over ``need'' based exchanges. This means that the agents 
rapidly distribute their holdings in an efficient way, thus allowing
money exchange to become the dominating mode of transaction. 

\item{}
Over long periods, one particular good is considered valuable by a
majority of the agents.  The monetary value of a good $j$ was defined as
the number of agents $M=M(j)$ that considered that particular good to be
the easiest to trade. Thus monetary value is a measure of consensus
among the agents in the system.

\item {}
Demand for a product, defined as the total number of times  $D$ that
product $j$ appears in the memory of the system,

\begin{equation}
D(j) \;=\; \sum_i T_{ij}\;,
\end{equation}

\noindent exhibits fluctuations which may be characterized by a Hurst
exponent $H \approx 0.7$ over a wide range of model parameters.
In contrast, the short and  long time statistics of the monetary
value, $M$ cannot be described by the same Hurst exponent.
\end{itemize}

In \cite{DS00} it was also discussed the appearance of absorbing
states for different products. Such states are reached when all agents 
have in stock a given product, and, at the same time, that product has
disappeared from the memory of all agents.

To simplify the model, we will initially assume that the production 
rate for all products is the same, and furthermore that production and
consumption are such that their combined stocks are constant.
This is achieved by requiring that every time an agent produces an unit
of some good, another unit is consumed and thus removed from its stock.
The production/consumption rate is given in terms of a parameter $p$,
that measures the probability that such events take place at a given
time step in the simulation. We have implemented this by imposing that
at each time step one of the agents has a probability $p$, of producing
and consuming goods. While the good being consumed by an agent is taken
at random among those in its stock, we have considered that it produces
a good with a probability proportional to its subjective desirability,
{\it i.e.} the times that item appears in its memory list.
This provides the classical negative feedback between abundance and value.

The results presented in the following section are independent of
details of the model. For example, one could consider production only
after encounters that lead to transactions, or after encounters that
lead to no transactions, and the conclusions reached would be robust
with respect to these modifications.

\section {Results.}

The described model contains four basic features, namely exchange of
information, trade associated to perception of demand, trade due to
fulfillment of need and finally production and consumption.
The fact that demand is taken into account tends to unfreeze the economy.
Products that by accident have become well distributed (present in all
agents' stocks), but which still occupy part of their memories, are still
traded and therefore redistributed in such a way that new need is generated.
On the other hand, adding need to an economy driven purely by perception
of value (demand) limits the dominance that one product would have in an
economy governed only by demand.
In other words, as the demand for a product grows, its rapid exchange
for other products naturally generates need, and thereby demand, for
other products.

If no other ingredients are added, products that both become well
distributed and forgotten by all agents will never be traded again.
As seen in Fig.~1, this increase in the number of forgotten
products ultimately leads to an economy where only a few products
are left for active trading. 
We see in the same figure how a small production/consumption term may
alleviate the absorbing state of a pure trading economy, thereby allowing
the system to settle into states where, at all times, a fraction of the
products are active, and all products become active at one time or another.

This means that the inclusion of production/consumption adds an external
source to an otherwise decaying system, thereby pushing it to a
non-equilibrium steady state where various products alternate being the
most demanded one. 
Although in our model production/consumption is not responsible for the
emergence of cooperativity and perception of value, in the long term it
is an essential ingredient for keeping the system alive and able to change.

When analyzing the number of units of a given product as a function of
time, and compare it with the demand for the same product, we notice that
a clear correlation appears between amount in stock and demand. 
As illustrated in Fig.~2, typically a product that is in low stock raises
rapidly in demand, and, consequently, is produced at a fast rate. 
Once it has become abundant, however, the demand for it rapidly
disappears. The number of units of the products decreases at a lower rate,
as it is removed from the stocks only when it is consumed, and, at any
given time, there is a much larger number of products in low demand than
in high demand.
We see in the same figure that the periods where there is a sizable
demand for a product coincide with those where the same product has a
high monetary value.

So far we have considered all products to be a priory equally easy or
difficult to produce, and even with this assumption obtained products,
that, for some time, become universally accepted means of exchange,
{\it i.e.} they take the role of money in the model system.
The model thus naturally enables us to examine relationships between
production and demand.
In Fig.~3 we show a simulation where one of the products
is a factor 50 more difficult to produce than all other products.
The upper panel of that figure shows that while the stock of one 
of the other products behaves in an essentially similar way as
in Fig.~2, the stock of the good having a low production rate
decreases gradually until it stabilizes at approximately 60\% of
the average value for all goods.
As observed in the lower panel of Fig.~3, the scarcity of the
difficult to produce good leads to a large increase in its
demand, which takes values much larger than for normal products.

A simple calculation allows to estimate the average steady state
demand for the scarce product, $D_s$, from the model parameters.
The probability that at a given time step the scarce product is
added to the memory is
\begin{equation}
\frac{D_s}{D_{Tot}}=
\frac{P_s(0)+D_s/D_{Tot}}{P_s(0)+(N_{pr}-1)\cdot P_n(0)+1}\,,
\label{demand}
\end{equation}

\noindent where $D_{Tot}=N_{ag}\cdot N_{mem}$ is the total memory of the
system and equals the maximum value of the demand for any given product,
and $P_s(0),P_n(0)$ are the probabilities that the scarce and the normal
product are absent from the selected agent's stock, respectively.
The numerator of the rhs of eq.~(\ref{demand}) represents the events
that the scarce product is added to the stock because of need or demand,
while in the denominator we have also included all other products' needs 
and demands.
Eq.~(\ref{demand}) may be simplified to
\begin{equation}
D_s=\frac{P_s(0)\cdot D_{Tot}}{P_s(0)+(N_{pr}-1)\cdot P_n(0)}\,,
\label{scarcedemand}
\end{equation}
which relates demand of the scarce product to how much
the product is needed relative to the other products.

\noindent In order to determine steady state values of $P_s(0),P_n(0)$
we have calculated the distribution of stocks by averaging their values
on the asymptotic region of the time evolution.
In Fig.~4 we show the probability distributions for a given agent
having $N$ units of either the scarce or a normal product, in the
stock of an agent. 
From it we find $P_s(0)\approx 0.34, P_n(0)\approx 0.01$;
by substitution of these values and the parameters of the simulation 
into eq.~(\ref{scarcedemand}) we obtain $D_s\approx 2000$, in 
agreement with the results shown in Fig.~3.
It is important to note that the value, $P_n(0)$, represents the 
typical need in the system, and thus determines the mode in which
the economy operates.
If it is large, the system is driven by need, and cooperativity
breaks down, and it is not possible for the system to maintain
sizable stocks of the scarce product.

It is interesting to remark that the probability distribution for
the stock of the scarce product, shown in Fig.~4 is adequately
described by a Poisson distribution with parameter $\lambda =1.2$.
This is consistent with the average stock of this product, also
observed in Fig.~3, $N_s\approx 60$, as one should expect that
$\lambda = N_s/N_{pr}$.
On the other hand, the distribution for the normal product is
clearly non-Poissonian, as the probability for having $N=0$ is
very depleted. 
The difference in these two distributions may be directly related 
to the high demand for the scarce product, which makes it the most 
frequently traded.

To show that this high trading activity is the essential feature
leading to the relatively high availability of the scarce product,
we show, in Fig.~5 the results of a simulation with the same
parameters of those of Fig.~3, but where trade was not allowed.
We see that, although the stock of an average product behaves about
the same as when trade is considered, its absence makes the stock
of the scarce product fall to very low levels. Therefore, production
alone is not sufficient to lead to a cooperative state where agents
have enough stock of scarce products. This can be achieved only
through trade.
In fact, we have numerically verified that both a relatively large
trade/production ratio, and certain abundance of products, measured
as the ratio $N_{unit}/N_{pr}$ is needed to maintain the cooperative
state.

As agents that interact are picked at random, the present model does not
consider spatial distribution of the agents.
However most of the presented results are robust to even the most restricted
spatial distribution: agents placed on a 1-d periodic lattice.
Fig.~6a shows the spatio-temporal development of agent consensus (monetary value)
of agents that consider a given product the most demanded one.
One observe that ``fashion" waves of this product appear and propagate
in the system. In Fig.~6b we show the corresponding overall behavior of demand,
value and total stock for parameters identical to the ones used in Fig.~2.
One observes persistent fluctuations in overall demand, that scale with a
Hurst exponent $H\sim 0.7$, as it was found in Ref.~\cite{DS00}. 
On the other hand, neither the consensus measure (the monetary value of
the product) nor its overall stock exhibit scaling. In the case of the stock,
the absence of scaling is noticed when one consider higher moments of
the fluctuations (not shown).
None of these quantities exhibits multiscaling.
Notice that the size of demand fluctuations does not depend of dimension,
whereas the consensus fluctuations are much smaller in the 1-dimensional case.

The quantitative comparison in Fig. 7a shows that spatial restrictions do not
change fluctuation sizes in demand and stock appreciably, but systematically
decrease the overall consensus, measured by the value of the product. 
Thus spatial localization increases the agreement between nearby agents,
but on the other hand makes fewer agents agree with each other.
In this way the system becomes fractionated in groups.
The typical size of the groups can measured through the
time-averaged spatial decay of demand overlap
$\langle T_{ik} T_{jk} \rangle=\sum_k T_{ik}T_{jk}$
with agent separation $x=j-i$, as plotted in Fig~7b.
Notice that a lower production/consumption rate $p$ increases correlation
lengths and overall correlations throughout the system.

\section{Conclusions and outlook}

We have earlier discussed \cite{DHSS00} how agents with memory trade to 
fulfill both their needs and  also to improve their future possibilities 
to trade.
In the present work we have focused on the feedback mechanism between
need and production of a product.
Overall production/consumption is not a needed element in building a
cooperative economy, but it is a necessary element in maintaining active
all products in the model economy.
Cooperativity appears as a consequence of trade and exchange of
information.

Key elements in maintaining the civilized/cooperative state is large memory,
enough richness ($N_{mem},N_{unit}$, respectively, in our model) and many 
trades/communications per production event ($p$ small enough).
For high cooperativity the system is able to maintain a rather large stock
of a product that is difficult to produce, simply because all agents develop
a large demand for this product.
Stock maintenance collapses when cooperativity between agents disappears.
This resembles the transition from a cooperative/civilized state with well
tempered amounts of all products and a disorganized state where some
products essentially disappear.

In summa, the introduction of production, and subsequent spatial localization
opens for studies of consequences of collapse of cooperativity on stock maintenance,
and of self organized groups of agents that internally agree on values,
but externally disagree on the value assessment with other groups.
\vspace{0.5cm}

R.D. thanks Nordita and the Niels Bohr Institute for hospitality
and support during his stay in Copenhagen during which this work
was performed, and acknowledges support from the Brazilian National
Research Council (CNPq).

\vspace{0.5cm}
{\bf Figure Captions}

\begin{itemize}

\item
Fig. 1. Number of products in the absorbing state for two values
of the production/consumption parameter $p$. This simulation was
performed for $N_{ag}=50$ agents, $N_{pr}=50$ products, $N_{unit}=100$
units of each product and a memory length $N_{mem}=100$. 
Time is defined as number of trades per agent.
See text for other details.

\item
Fig. 2. Demand, value and stock for a typical product, as a function
of time, in a simulation with the same parameters as Fig. 1.
Here $p=0.02$ for all products. 

\item
Fig. 3. 
\noindent Upper panel: time evolution of the stocks of the scarce (low $p$)
and a normal (high $p$) product for a system with the same parameters as in
the previous figures. Normal products have $p=0.02$, and the scarce product
had its $p$ parameter value reduced by an additional factor $r=0.02$.
See text for further details.

\noindent Lower panel: time evolution of the demand for the two
products illustrated in the upper panel.

\item
Fig.~4. Probability distributions for a given agent having $N$ units,
of either the scarce or a normal product, in its stock. Same parameter
values as in Fig.~3.

\item
Fig.~5. Similar to the upper panel of Fig.~3, but without trading of goods.

\item
Fig.~6.
\noindent Upper panel: Spatio-temporal distribution of value regarding
a particular product in the 1-d version of the model.

\noindent Lower panel: Overall variability in total demand, value and stock
of the same product as in the upper panel.
All parameters are identical to those of Fig. 2, thus allowing a direct assessment
of the influence of the 1-d geometry.

\item
Fig.~7. 
\noindent Upper panel: Fluctuation analysis of a particular product in 
the random neighbor and the 1-d versions of the model. 
The model parameters are the same that in the previous figures. 
See text for details.

\noindent Lower panel: Spatial correlations in the  1-d version of the model, 
illustrating that local demand agreement is larger in the 1-d case,
although overall agreement is similar in the two cases.

\end{itemize} 

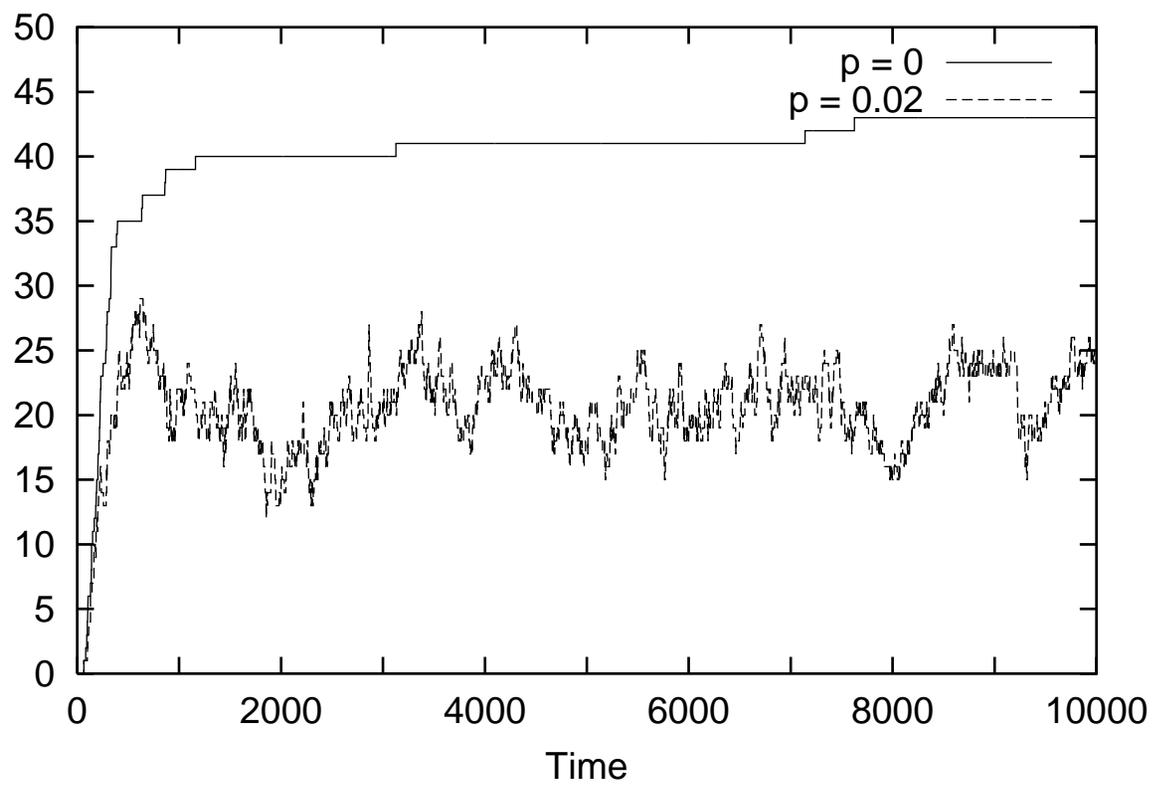
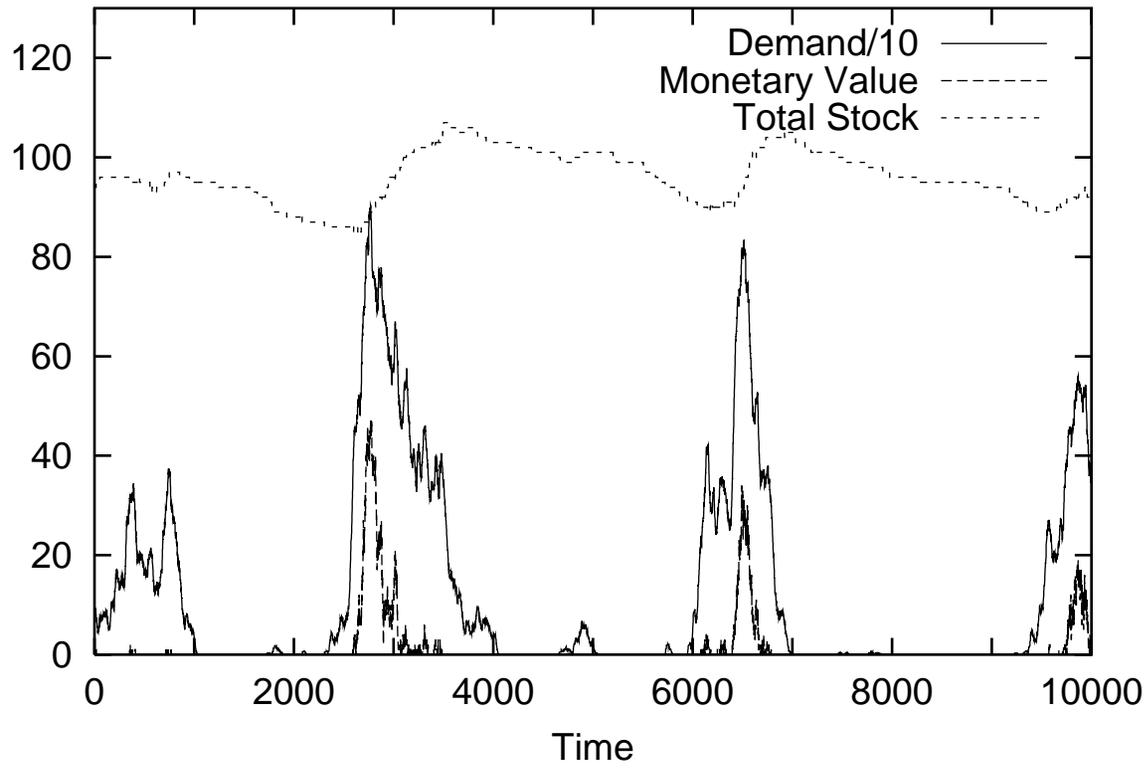
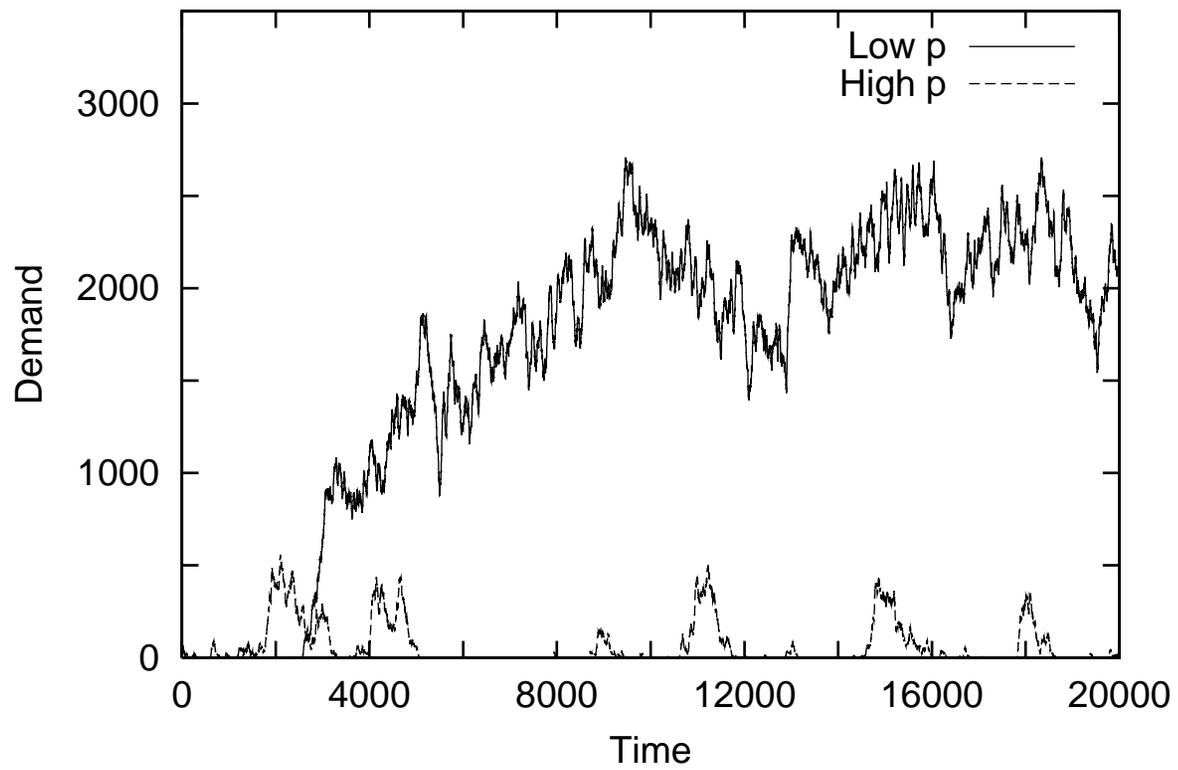
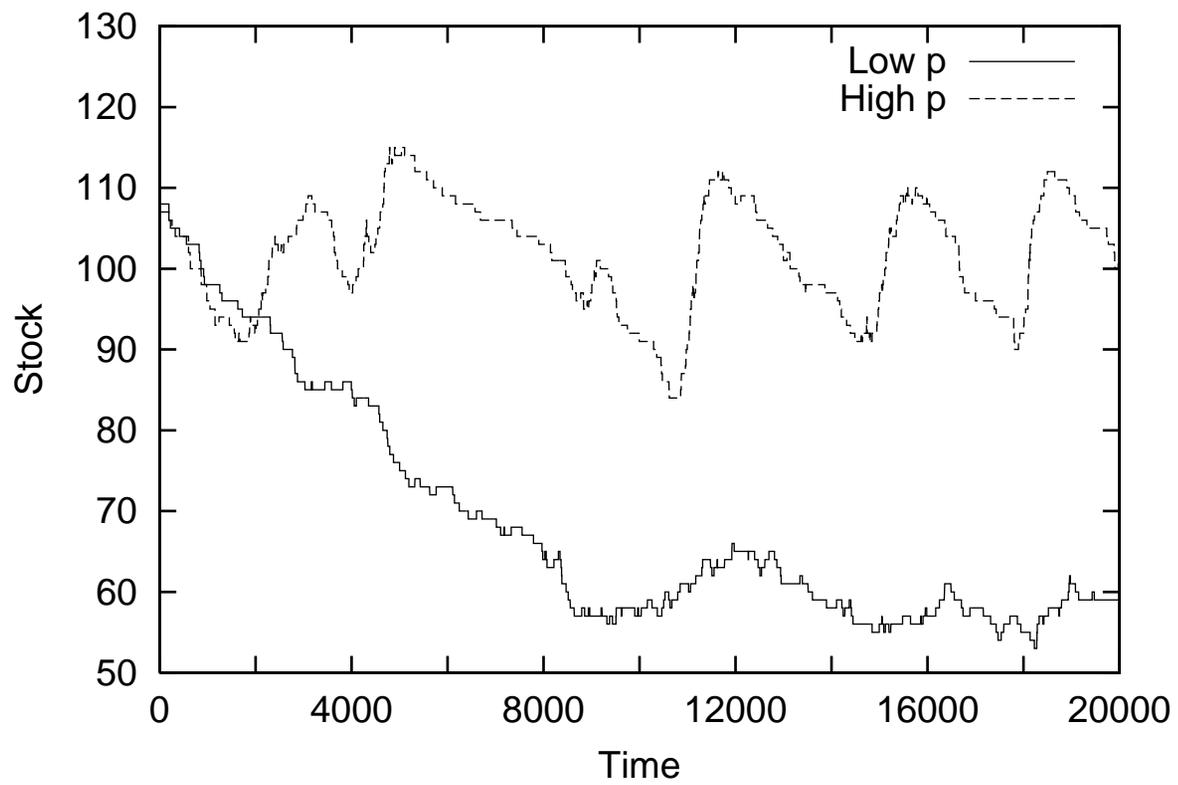
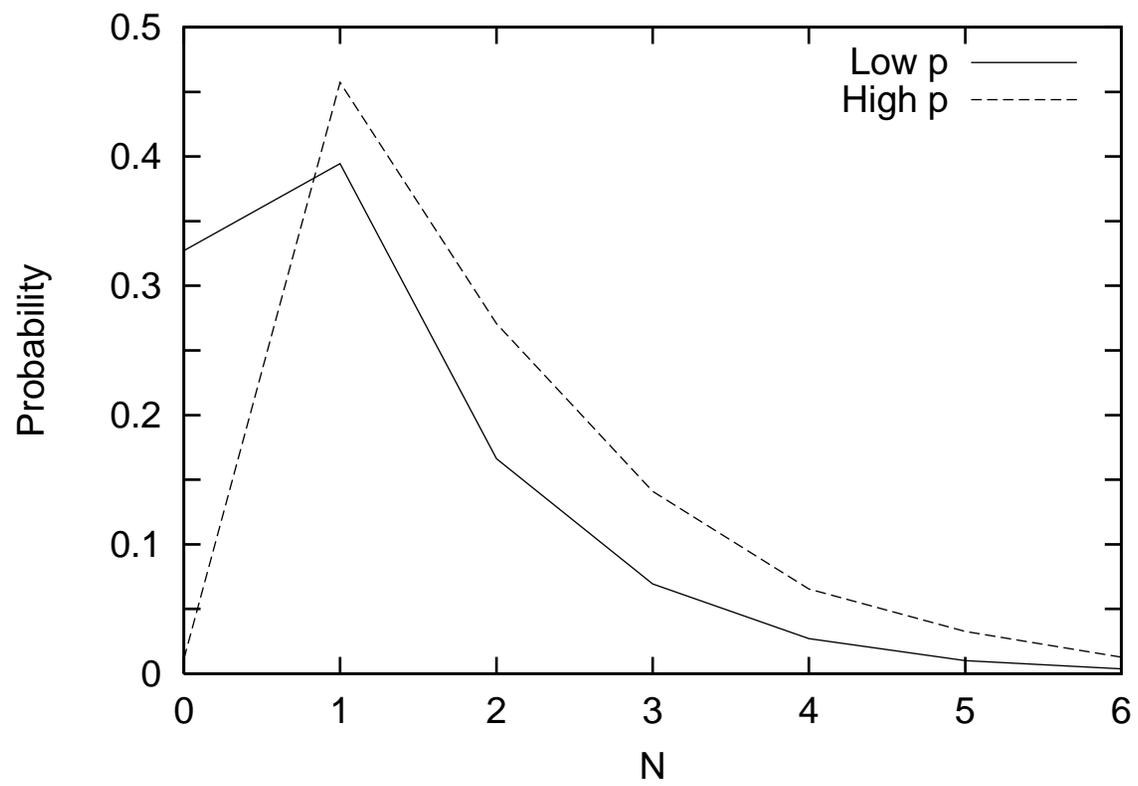
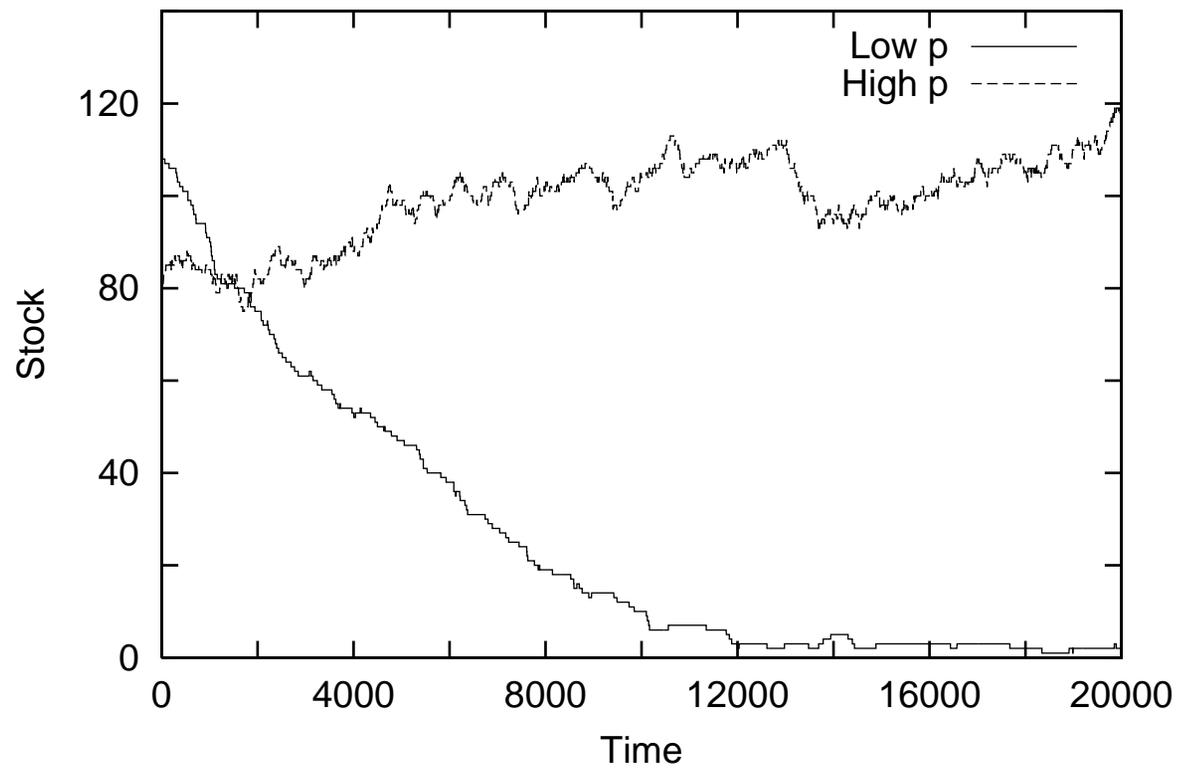
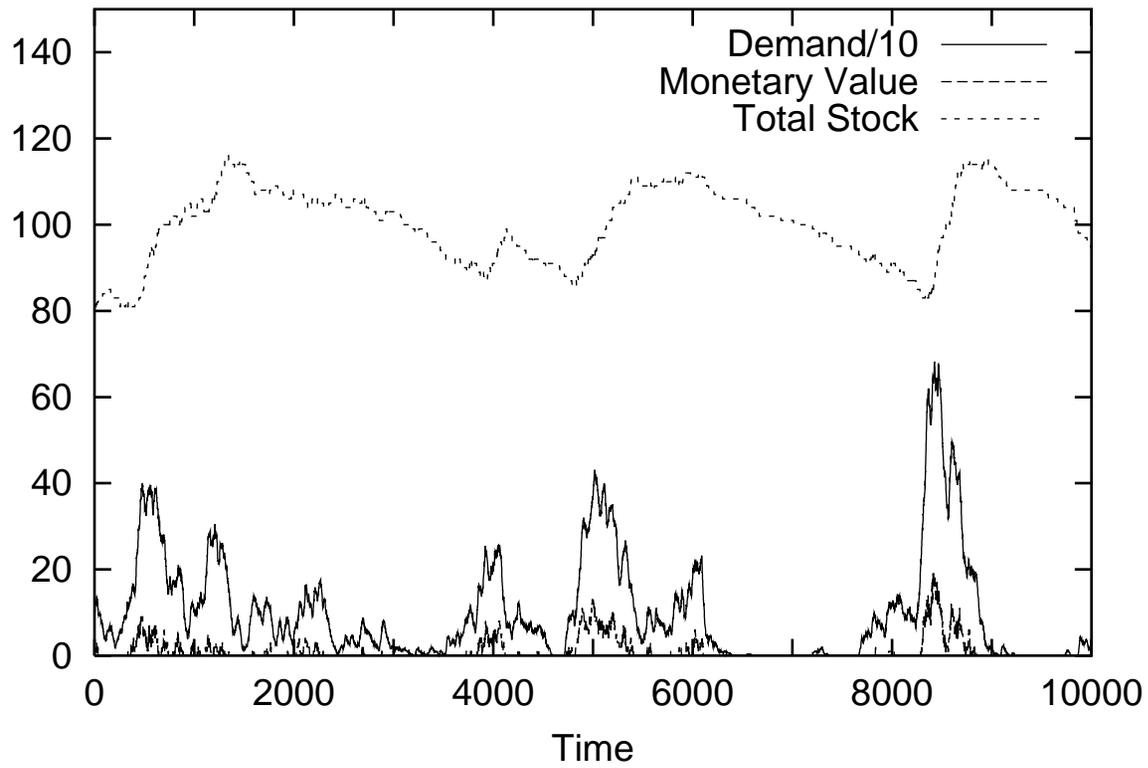
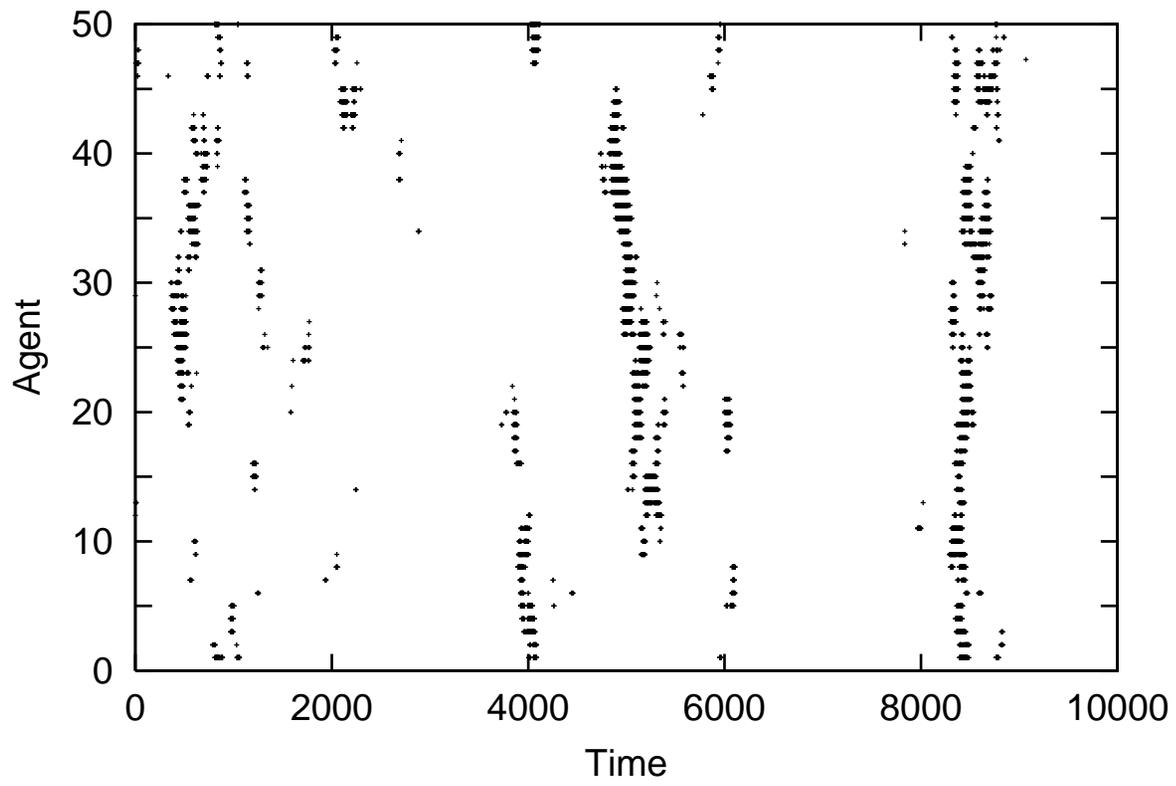
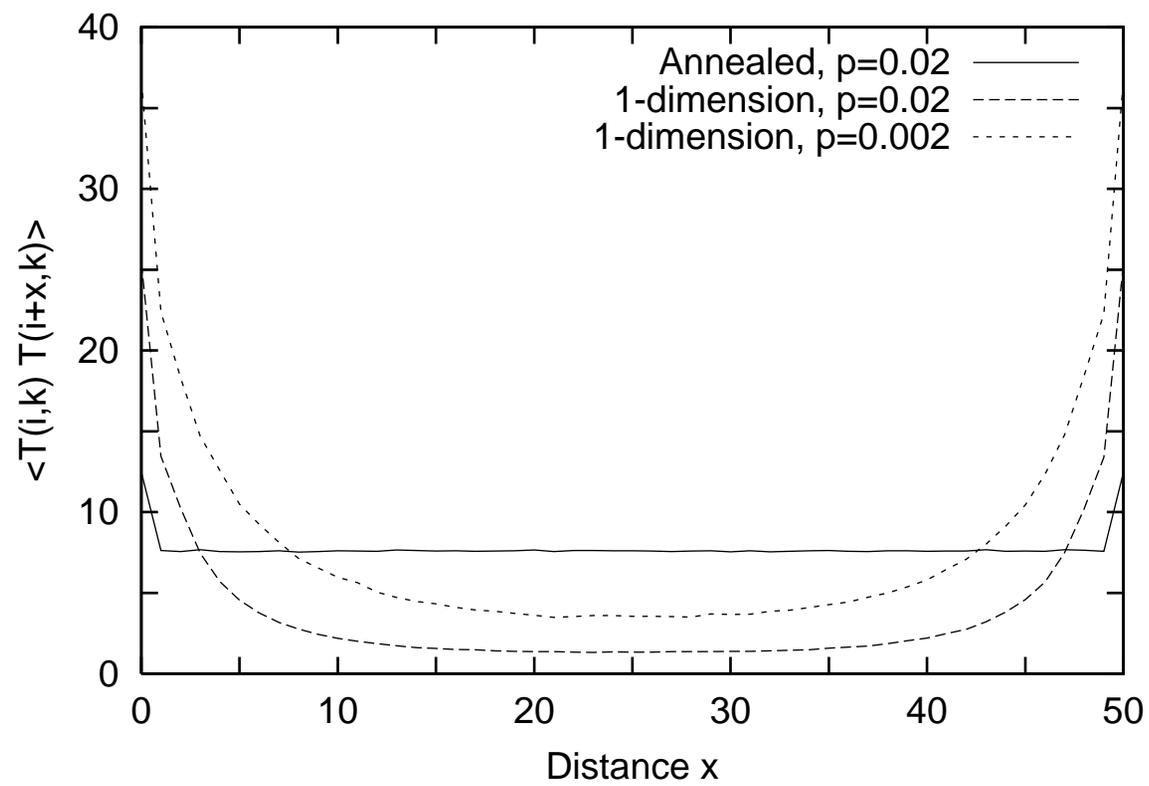
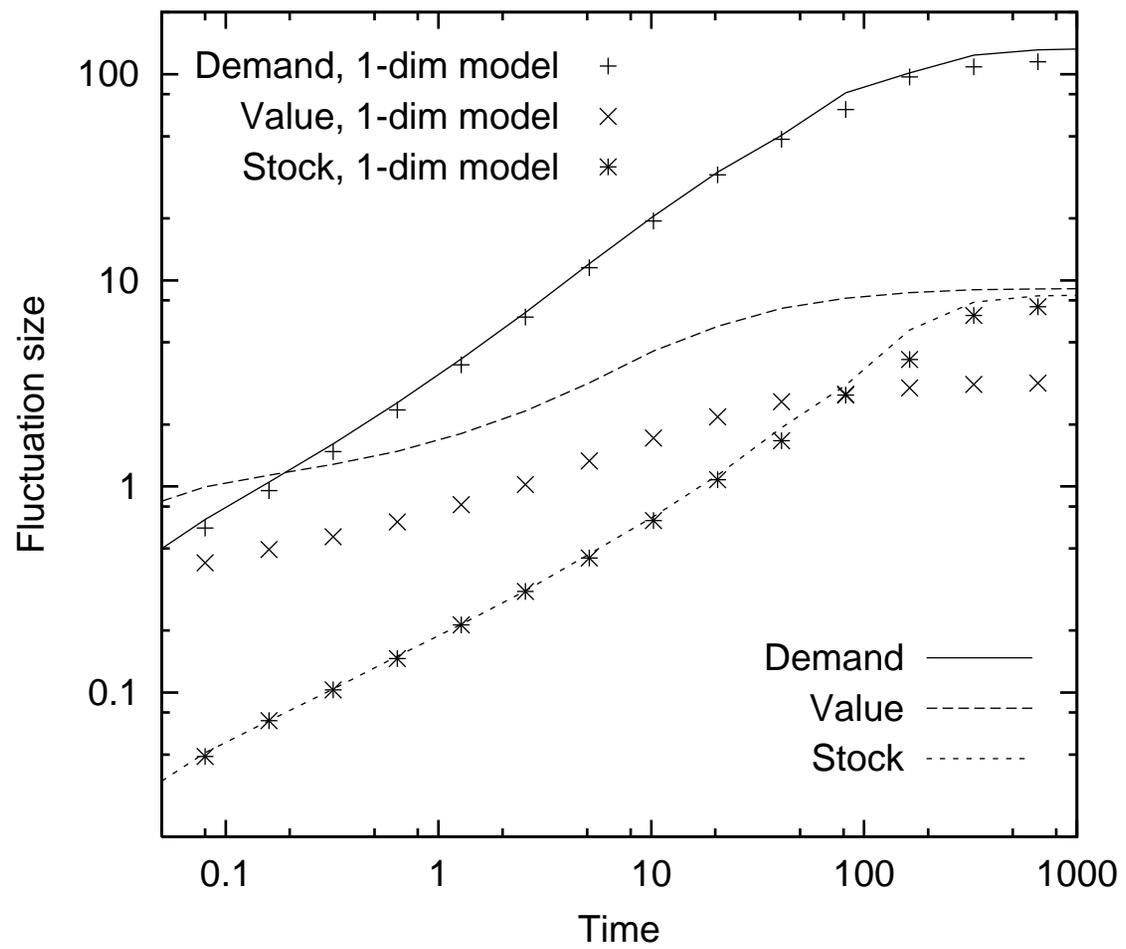
\end{document}